\documentclass{article}

\usepackage[utf8]{inputenc}
\usepackage[T1]{fontenc}
\usepackage{hyperref}
\usepackage{url}
\usepackage{booktabs}
\usepackage{amsfonts}
\usepackage{nicefrac}
\usepackage{microtype}
\usepackage{graphicx}
\usepackage{natbib}
\usepackage{doi}

\title{On the Security Implications of PQC in TLS: Handshake Exhaustion and IDS Degradation}

\author{
Lin-Fa Lee\\
Department of Institute of Artificial Intelligence Innovation\\
National Yang Ming Chiao Tung University\\
Hsinchu, Taiwan\\
\texttt{prologue.ii14@nycu.edu.tw}
\and
Yi-Yu Chang\\
Department of Institute of Artificial Intelligence Innovation\\
National Yang Ming Chiao Tung University\\
Hsinchu, Taiwan\\
\texttt{daniel282907@gmail.com}
\and
Chia-Mu Yu\\
Department of Institute of Electrical and Computer Engineering\\
National Yang Ming Chiao Tung University\\
Hsinchu, Taiwan\\
\texttt{chiamuyu@gmail.com}
\and
Kuo-Hui Yeh\\
Department of Institute of Artificial Intelligence Innovation\\
National Yang Ming Chiao Tung University\\
Hsinchu, Taiwan\\
\texttt{khyeh@nycu.edu.tw}
}

\hypersetup{
pdftitle={On the Security Implications of PQC in TLS: Handshake Exhaustion and IDS Degradation},
pdfauthor={Lin-Fa Lee, Yi-Yu Chang, Chia-Mu Yu, Kuo-Hui Yeh},
pdfkeywords={Post-Quantum Cryptography, TLS, DDoS, IDS}
}

\begin{document}

\maketitle

\begin{abstract}

Post-Quantum Cryptography (PQC) is increasingly being integrated into TLS 1.3 to enhance resilience against quantum enabled attacks. However, the additional computational and communication overhead introduced by PQC primitives during the handshake phase may also amplify the impact of TLS handshake exhaustion attacks, leading to more severe Distributed Denial-of-Service (DDoS) threats.

In this study, we establish an empirical testbed consisting of one PQC enabled TLS server and ten attacking nodes, generating over 16.5 GB of mixed traffic data that includes both legitimate browsing behavior and high intensity handshake exhaustion attacks.

Experimental results show that PQC-TLS can prolong periods of sustained high CPU utilization on the server by up to 88 times, significantly amplifying the effectiveness of such attacks. Furthermore, we evaluate state-of-the-art deep learning-based Intrusion Detection Systems (IDS) and observe a substantial decline in attack detection performance under PQC traffic conditions.

In particular, exosphere achieves only around 50\% recall, while HyperVision’s AU-ROC degrades to near random levels (0.49), revealing critical detection blind spots in existing IDS when operating in PQC environments.

The main contributions of this work are threefold: (1) we systematically quantify and analyze the root causes of IDS detection blind spots in PQC settings; (2) we publicly release a comprehensive PQC-DDoS hybrid traffic dataset, including precise attack timestamps and server side resource monitoring data; and (3) we open source all experimental code and AWS deployment scripts, enabling a fully reproducible cloud based testing environment.

These resources aim to support both academia and industry in developing next generation PQC aware intrusion detection systems.

\end{abstract}


\section{INTRODUCTION}

Today, a wide range of cryptographic algorithms and digital signature schemes are used to ensure the confidentiality, integrity, and availability of data during transmission and storage. In most traditional cryptosystems, security is not based on the assumption that “attacks are theoretically impossible,” but rather on the premise that “attacks are computationally infeasible given current computational models and available resources”~\cite{xiao2019edge}. In other words, even if a potential attacker knows a viable strategy, the required computational time, memory, or energy cost grows exponentially with the security parameters, making it practically impossible to succeed within an acceptable timeframe. This type of security is commonly referred to as computational security~\cite{liu2020security}.

However, certain quantum algorithms, such as Shor’s and Grover’s algorithms, can significantly reduce the computational complexity of some underlying hard problems~\cite{piastou2024investigating}. As a result, once quantum hardware reaches a critical scale, cryptosystems that previously relied on these hard problems could shift from being practically secure to being practically vulnerable~\cite{joseph2022transitioning}.

This shift has led to the development of Post-Quantum Cryptographic (PQC) algorithms that are resilient against sufficiently large and fault-tolerant quantum computers. These algorithms must not only withstand quantum attacks but also remain compatible with implementations and operations on classical computers~\cite{wang2025review}.

To support this transition, the National Institute of Standards and Technology (NIST) launched an open initiative aimed at standardizing public-key cryptographic algorithms that can be deployed in practice and resist both classical and quantum attacks, serving as replacements for traditional mechanisms like RSA and ECC~\cite{truong2026nist}.

PQC is transitioning from theoretical research to practical deployment. In 2022, NIST selected CRYSTALS-Kyber for key encapsulation, along with three digital signature schemes—CRYSTALS-Dilithium, FALCON, and SPHINCS+—as the core of the future standard~\cite{truong2026nist}. Following this, companies have begun integrating PQC into everyday products. For example, Chrome now supports a new hybrid key exchange group: initially X25519Kyber768, and later, after Kyber standardization, it aligned naming and versions to X25519MLKEM768~\cite{mrinal2025performance}. Cloudflare also provides PQC/hybrid support for origin servers, enabling end-to-end protection, specifically on the Cloudflare-to-origin segment, through hybrid key exchange based on ML-KEM~\cite{egbuagha2025post}.

However, the practical deployment of PQC introduces additional latency and communication overhead due to its inherent characteristics~\cite{almutairi2025resilience}. Compared with traditional ECDHE/ECDSA, mechanisms such as ML-KEM increase both the computational burden and packet size during the TLS handshake, which in turn raises latency and bandwidth consumption~\cite{shahid2026post}. In our measurement environment, when TLS 1.3 performs a hybrid key exchange using \{X25519+ML-KEM-768\}, the server experiences significantly higher computational costs for each complete handshake, and the associated communication overhead is also substantially increased.

These increased costs also exacerbate the existing resource asymmetry in the TLS handshake phase, making handshake exhaustion attacks more threatening in the context of PQC. During the handshake, the server typically performs expensive cryptographic operations before completing application-layer authentication. This allows an attacker to trigger a large number of handshake attempts, consuming server CPU and bandwidth resources at a relatively low network cost~\cite{rescorla2002sslacc}.

When PQC amplifies both the computational and transmission costs of each handshake by multiple times, attackers can inflict much higher resource consumption on the server with the same input traffic. This reduces the number of legitimate connections the system can handle and increases the risk of denial-of-service. In this work, we quantify the amplification effect of handshake exhaustion under PQC-TLS and provide empirical evidence to support it.

Traditionally, TLS handshake exhaustion has been mitigated through protocol-level mechanisms and IDS-based detection. TLS 1.3 introduces measures to reduce cost asymmetry, allowing the server to request clients to prove address reachability via cookies, providing a degree of DoS protection. Cookies also support a more “stateless” HelloRetryRequest, reducing the amount of state the server must maintain during the early handshake~\cite{maehren2025towards}. IDS systems, on the other hand, rely on observable features such as packet size and connection behavior from the same source IP to detect anomalies~\cite{thottan2003anomaly}.

However, with the advent of PQC-TLS, no prior studies have evaluated the detection effectiveness in a post-quantum context. This gap has led to renewed attention to previously underexplored attack vectors, creating a significant security blind spot. To evaluate IDS performance under PQC-TLS, we built an empirical testbed comprising one PQC-TLS server implementing ML-KEM-768 and ML-DSA-65, and ten attack nodes, generating over 16.5 GB of mixed traffic that includes legitimate browsing behaviors and high-intensity TLS handshake exhaustion attacks.

We assessed two mainstream deep learning-based detection systems and compared their performance under both traditional TLS and PQC-TLS environments to establish a baseline. The experimental results confirmed our expectations: both HyperVision and Exosphere exhibited substantially degraded detection performance under PQC-TLS compared to their performance against other attack types.

We summarize our findings as follows.First, PQC significantly amplifies the resource impact of TLS handshake exhaustion attacks. Under the same attack intensity, PQC-TLS prolongs the duration of high CPU utilization on the server by up to $88\times$ compared to traditional TLS, suggesting that the additional computational overhead of PQC increases the effectiveness of such attacks.

Second, existing machine learning--based IDS solutions show limited capability in detecting TLS handshake exhaustion attacks. Exosphere achieves a recall of approximately 50\% under both TLS configurations, which is effectively comparable to random guessing. Similarly, HyperVision yields an AUROC of 0.49, again comparable to random performance, suggesting that graph-based analysis fails to capture the underlying attack patterns.

A key factor behind this failure is a lack of representative training data. In traditional TLS settings, handshake exhaustion has been considered a relatively minor threat and is therefore largely absent from existing benchmarks. However, with the adoption of PQC, not only is this attack amplified into a practical threat, but its distinct traffic characteristics further degrade IDS detection performance—creating a dual security blind spot.

We make the following contributions.First, through a real-world testbed and realistic attack simulations, we demonstrate that the computational overhead introduced by PQC substantially amplifies the impact of TLS handshake exhaustion attacks. Under identical attack intensity, the duration of sustained high CPU load on the server increases by $88\times$, providing concrete evidence of this amplification effect.

Second, we release an open dataset and reproducible environment. Our dataset contains over 16.5 GB of mixed traffic, including both traditional TLS and PQC-TLS scenarios, with full packet captures, precise attack annotations, and server-side resource monitoring data. This dataset fills a critical gap in PQC-related attack traffic. In addition, we open-source all experimental code and AWS deployment scripts, enabling a fully reproducible cloud-based testing environment and lowering the barrier for future research.
\section{RELATED WORK}

This section reviews the critical literature across the domains of post-quantum cryptographic (PQC) performance, advancements in DDoS and handshake exhaustion detection, and the current landscape of benchmark datasets in top-tier security venues.

\subsection{PQC Security Analysis}

The migration toward quantum-resistant communication has been accelerated by the NIST PQC standardization project, which recently finalized the first core standards in August 2024: FIPS 203 (ML-KEM), FIPS 204 (ML-DSA), and FIPS 205 (SLH-DSA). While these algorithms are cryptographically robust, their integration into real-world network protocols introduces unprecedented performance and resource trade-offs.

\begin{itemize}
\item \textbf{Deployment in Network Protocols} \\
Research on integrating PQC into TLS 1.3 and SSH has highlighted the challenges of “handshake inflation.” Sikeridis et al. conducted early large-scale performance evaluations, demonstrating that adopting PQC signature schemes such as Dilithium (ML-DSA) and Falcon can increase handshake latency by up to 300\%, depending on network volatility~\cite{thottan2003anomaly}.

Recent security work has increasingly focused on defending against HNDL-style attacks by adopting hybrid key exchange designs that combine ECDH with ML-KEM. As shown by Sikeridis et al., performance evaluations of post-quantum TLS 1.3 implementations suggest that these hybrid approaches can maintain interoperability while still providing fallback security. Meanwhile, lattice-based signature schemes such as Dilithium seem to be a practical choice for authentication, as they do not noticeably affect connection setup time~\cite{sikeridis2020post}.

Furthermore, architectures like KEMTLS aim to replace signatures with KEM-based authentication to reduce the bandwidth burden, though they often require additional round trips (RTTs)~\cite{schwabe2020post}.\\

\item \textbf{Performance Overhead and Hardware Constraints} \\

The primary bottleneck in PQC handshakes is substantial data inflation. ML-KEM and ML-DSA introduce approximately 10 KB of additional data into negotiation messages compared to the few hundred bytes required for classical ECC. Kampanakis et al. observed that this inflation often exceeds the default TCP initial congestion window of 10 Maximum Segment Sizes (MSS), triggering additional RTTs that double latency in high-delay environments~\cite{schwabe2020post}. While computational costs for KEM operations can be optimized using Intel AVX-512 extensions to achieve a 1.64× speedup, the memory requirements for signature verification in SPHINCS+ remain prohibitive for resource-constrained IoT devices. NVIDIA has also introduced cuPQC to explore GPU-accelerated PQC, achieving 13.3 million key generations per second, highlighting the gap between hardware-accelerated servers and unoptimized clients.
\end{itemize}
\subsection{DDoS Detection and Handshake Exhaustion}

The increased complexity of PQC handshakes has resurrected concerns regarding resource asymmetry, where a client forces a server to perform expensive operations with minimal effort.
\begin{itemize}
\item \textbf{Handshake Exhaustion Research} \\

Traditional handshake exhaustion has evolved from simple SYN floods to sophisticated application-layer exploits. A landmark study by Shi et al. introduced the “X.509DoS” class of attacks, identifying 18 new zero-day vulnerabilities in cryptographic libraries such as OpenSSL and GnuTLS~\cite{kampanakis2024impact}. Their research shows that crafted X.509 certificates can exploit implementation flaws in ASN.1 parsing or mathematical modules to cause sustained CPU saturation (>80\%) or memory exhaustion before any signature verification occurs. In a PQC context, the higher dimensionality of lattice-based arithmetic provides more opportunities for such resource-heavy loops, intensifying the risk of complete system unresponsiveness.\\

\item \textbf{ML-based IDS Research} \\

As encryption obscures traditional packet features, research has shifted toward machine learning-based IDS (ML-IDS). HyperVision proposed a real-time unsupervised graph learning method that detects abnormal interaction patterns by analyzing flow interaction graphs~\cite{shi2025x}. It achieves a detection throughput of 80.6 Gb/s by learning structural features rather than specific signatures.

Conversely, Exosphere utilizes deep semantic analysis of packet length patterns to identify tunneled flooding traffic, including stealthy DDoS attacks without accessing encrypted headers or payloads~\cite{fu2024detecting}.

Despite these advances, supervised models suffer from poor generalization when faced with unseen attack types, a challenge defined as the “Closed World” problem by Sommer and Paxson~\cite{sommer2010outside}.
\end{itemize}
\subsection{Network Traffic Datasets}

The efficacy of ML-based detection is fundamentally limited by the representativeness and quality of training data.
\begin{itemize}
\item \textbf{Limitations of Existing Datasets} \\

Public benchmarks such as CIC-IDS2017 and CIC-DDoS2019 are current gold standards but have been shown to contain documented errors. Troubleshooting by Rosay et al.~\cite{rosay2022network} revealed that over 20\% of traffic traces in CIC-IDS2017 were mislabeled or incorrectly reconstructed due to artifacts in feature extraction tools.

Furthermore, many datasets are generated in synthetic laboratory settings that fail to capture the “base rate fallacy” and concept drift characteristic of real-world production networks. Model accuracy often drops from 99\% on synthetic data to 92\% when applied to real-world traffic, indicating a significant “laboratory success gap.”\\

\item \textbf{The Need for PQC-specific Datasets} \\

PQC migration fundamentally alters the statistical profile of “benign” network traffic. Handshakes now feature 7× larger message sizes, different fragmentation patterns, and increased endpoint processing times~\cite{kampanakis2024impact}.

Current benchmarks model benign traffic based on RSA/ECC signatures of a few hundred bytes; without PQC-specific datasets, IDS models will likely yield high false positive rates by misidentifying legitimate post-quantum negotiations as anomalous floods or DDoS attacks. Consequently, constructing datasets that reflect diverse PQC behaviors is essential for developing robust, quantum-ready defense mechanisms.
\end{itemize}

\section{ATTACK SCENARIO}

We consider a typical TLS handshake exhaustion attack scenario. The attacker provisions multiple distributed nodes using virtual machines and launches a large number of handshake requests toward a target PQC-TLS server. The goal is to exhaust server-side CPU resources, thereby preventing the system from serving legitimate users.

Notably, the attacker does not need to break the underlying PQC primitives or compromise the server’s private key. Instead, the attack exploits the inherent computational asymmetry of the TLS handshake. While the client only initiates low-cost handshake attempts, the server is required to perform expensive cryptographic operations, including KEM decapsulation, digital signature verification, and certificate chain processing.

By rapidly establishing and terminating connections or aborting them before the handshake completes, the attacker can force the server to repeatedly execute these costly operations. In the PQC setting, this effect is further amplified, as the computational complexity of ML-KEM and ML-DSA is significantly higher than that of traditional ECDHE and ECDSA.

We assume that basic network-layer filtering mechanisms are already in place. However, attack traffic can still bypass these defenses, for example, by distributing requests across multiple source IPs and maintaining seemingly reasonable request rates. As a result, detection must rely on application-layer mechanisms such as IDS.

Based on this threat model, we investigate three key research questions:

\begin{itemize}
\item \textbf{RQ1: How does PQC affect the impact of TLS handshake exhaustion attacks?} \\
We quantify the differences in server-side resource consumption under identical attack intensity in PQC-TLS and traditional TLS settings. Specifically, we compare CPU utilization, the duration of sustained high load, and the accumulation of TCP connections across the two configurations.

\item \textbf{RQ2: Can existing ML-based IDS effectively detect attacks in a PQC environment?} \\
We evaluate two representative deep learning-based detection systems, Exosphere and HyperVision, and measure their performance on mixed traffic containing both PQC-TLS and traditional TLS flows.

\item \textbf{RQ3: If detection fails, what are the root causes?} \\
We investigate whether detection failures stem from PQC-specific traffic characteristics or from other factors, such as the composition of training datasets or feature engineering choices. This analysis is conducted by comparing detection performance across TLS configurations and examining differences in traffic features.
\end{itemize}
\section{EXPERIMENTAL METHODOLOGY}

\subsection{Infrastructure Setup}

\begin{figure}
\centering
\includegraphics[width=\textwidth]{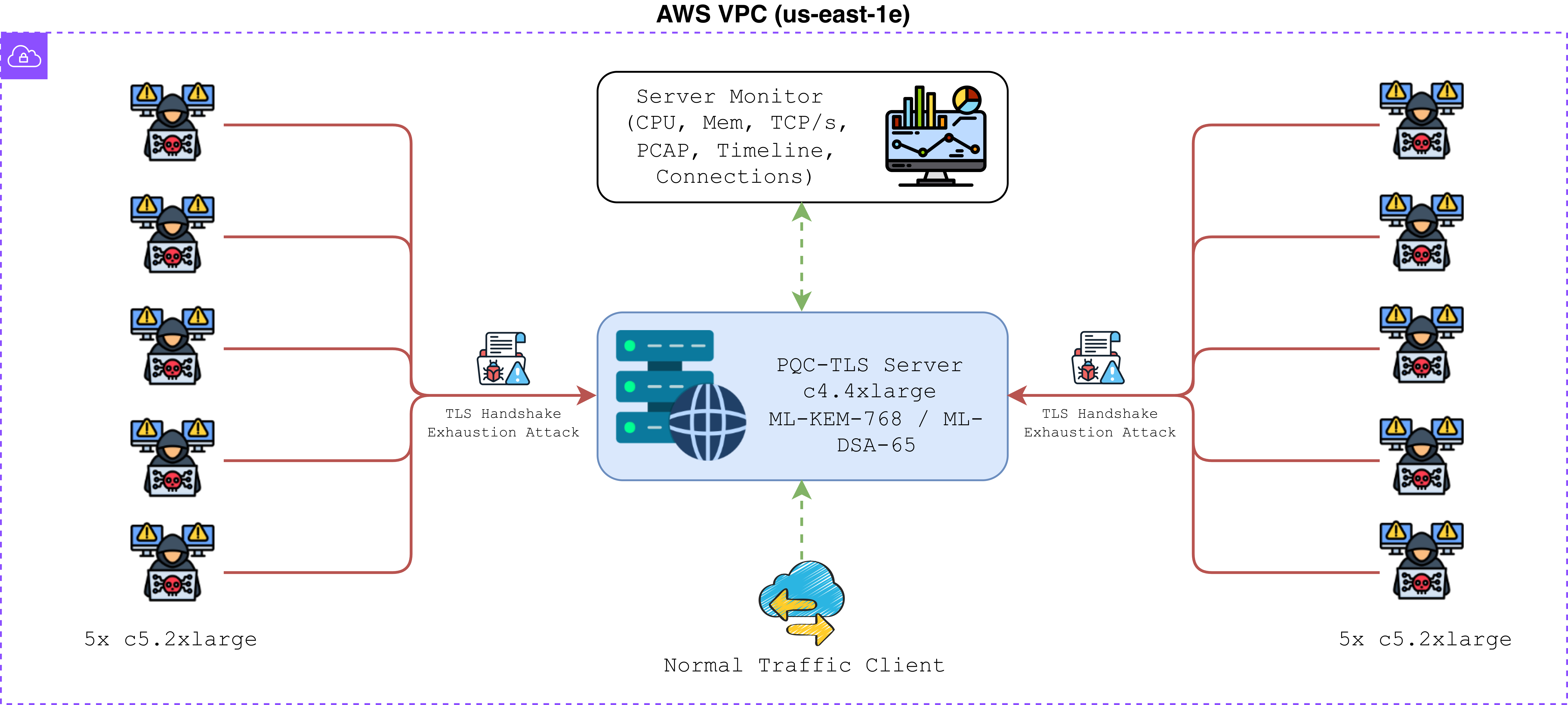}
\caption{AWS Based Experimental Infrastructure (1 PQC-TLS Server + 10 Attack Nodes)}
\label{fig1}
\end{figure}

We deploy our experimental environment on Amazon Web Services (AWS), consisting of one server and ten attack nodes, as illustrated in Fig.~\ref{fig1}. All instances are located within the same AWS region to minimize network latency.

The server runs on an EC2 c4.4xlarge instance, equipped with 16 vCPUs and 30 GB of RAM. Each attack node is provisioned as a c5.2xlarge instance with 8 vCPUs and 16 GB of RAM. All nodes communicate with the server through the internal AWS VPC network.The server is configured with two TLS setups for comparison. The PQC-TLS configuration is based on the OpenQuantumSafe nginx image and is deployed using Docker containers. This setup integrates ML-KEM-768 for key exchange and ML-DSA-65 for digital signatures and certificate issuance. The TLS handshake adopts a hybrid key exchange mechanism that combines traditional X25519 with ML-KEM-768, providing quantum resistance while maintaining backward compatibility.

For the baseline, the traditional TLS configuration uses ECDSA P-256 with X25519 key exchange, also deployed in a separate Docker container.To quantify the impact of attacks on server resources, we implement a comprehensive monitoring framework. On the server side, automated scripts continuously record key system metrics, including CPU utilization, memory usage, TCP connection counts, overall system load, and timestamps.

On the attacker side, each node logs the precise start time, end time, duration, and total number of connections for every attack instance. These logs are later used to construct ground truth labels.In addition, the server captures full network traffic using tcpdump, storing all packets in PCAP files covering the TLS ports. These traces are used for subsequent IDS evaluation. All monitoring and traffic capture mechanisms are activated prior to each experiment to ensure full coverage.

\subsection{Traffic Generation}

To construct a realistic mixed traffic environment, we generate both benign browsing traffic and attack traffic concurrently. Benign traffic is produced using a dedicated script that simulates legitimate user behavior. The script sends HTTPS requests to seven different endpoints on the server and rotates common browser User-Agent headers to increase diversity. 

The inter-arrival time between requests is randomly sampled between 1 and 10 seconds, resulting in an average rate of approximately 1.3 requests per second, which reflects a natural browsing pattern of a single user. This benign traffic runs continuously throughout the experiment, serving as a background baseline. Attack traffic follows a TLS handshake exhaustion strategy. We deploy attack scripts across ten distributed AWS EC2 nodes, each continuously initiating TLS handshake requests using OpenSSL.

For the traditional TLS setup, the attack consists of standard TLS 1.3 connection attempts. In the PQC-TLS setting, the attack explicitly specifies the use of the ML-KEM-768 key exchange group. Each attack connection is terminated immediately after the handshake is initiated, forcing the server to repeatedly perform expensive handshake computations without progressing to the data transmission phase. Each attack round lasts approximately 2–3 minutes, generating 10–15 million connection attempts, followed by a pause of 1–5 minutes before the next round begins. In total, 14 independent attack rounds are conducted over an experimental duration of approximately 60 minutes.

For each attack, we record precise start and end times, duration, and total connection counts. These records are later used to construct ground truth labels for IDS evaluation. This setup reflects a realistic scenario in which benign and malicious traffic coexist, and attacks occur in bursts rather than continuously, mirroring real-world DDoS behavior. Importantly, the traffic generation parameters are identical for both traditional TLS and PQC-TLS environments, with the only difference being the cryptographic algorithms used during the handshake, ensuring a fair comparison.

\subsection{Dataset Composition}

The experiment produces two complete mixed traffic PCAP datasets. The traditional TLS dataset contains 31,453,919 packets, while the PQC-TLS dataset contains 17,081,216 packets. The difference in packet volume is primarily due to slower handshake processing in PQC. Under the same attack intensity (approximately 12–13 million connection attempts per round), the PQC-TLS server completes fewer handshakes, resulting in fewer generated packets. We record precise timing information for all attack events, including start time, end time, duration, and connection counts. Based on these records, each packet is assigned a ground truth label: packets with timestamps falling within any attack interval are labeled as malicious, while all others are labeled as benign. Using this labeling method, the proportion of malicious packets reaches 99.998\% in both datasets. This extreme class imbalance is not an artifact of labeling, but rather a fundamental characteristic of TLS handshake exhaustion attacks. Each attack round generates 10–15 million connections within 2–3 minutes, resulting in approximately 169 million (traditional TLS) and 181 million (PQC-TLS) total connections across all rounds. In contrast, benign traffic operates at an average rate of 1.3 requests per second, producing only about 4,680 requests over 60 minutes.

As a result, attack traffic overwhelmingly dominates the dataset, reflecting the core strategy of distributed denial-of-service (DDoS) attacks: exhausting server resources through massive request volumes. This distribution closely mirrors real-world attack conditions and has important implications for the selection and interpretation of IDS evaluation metrics.

The datasets preserve the native characteristics of encrypted traffic. All TLS handshakes remain fully encrypted, and the PCAP files contain no decryption keys or plaintext content. This setup aligns with real-world IDS deployment, where detection systems can only access encrypted packets and must rely on metadata—such as packet size, timing, and flow-level features—rather than payload content. Finally, both datasets are generated under identical experimental conditions (same server specifications, attack node configuration, and attack intensity), with the only difference being the cryptographic algorithms used in the handshake, ensuring full comparability of results.

\subsection{Target Detection Systems}

We evaluate two representative IDS systems: Exosphere and HyperVision. These systems reflect two distinct design paradigms in modern ML-based intrusion detection and are both publicly available with pre-trained models, allowing us to assess their out-of-the-box detection performance. Exosphere adopts a supervised learning approach, leveraging a convolutional neural network (CNN) to analyze packet length sequences and extract deep semantic patterns for detecting flooding attacks hidden within encrypted tunnels. Its key advantage is that it does not require packet decryption; instead, it relies solely on temporal correlations in packet sizes.

In our evaluation, we do not perform any fine-tuning or re-training for TLS handshake exhaustion, in order to assess generalization capability to unseen attack types. We use the default configuration provided by Exosphere, including a batch size of 64 and a waterline threshold of 0.5. 

HyperVision, in contrast, adopts an unsupervised learning approach. It transforms network traffic into a flow interaction graph and applies graph neural networks (GNNs) to detect structural anomalies. Rather than relying on labeled data or specific attack patterns, HyperVision learns the structural characteristics of normal traffic and flags significant deviations as anomalies.

To improve scalability, HyperVision employs a short-flow aggregation mechanism that groups numerous short-lived connections into a smaller number of edges. The system outputs detection results in terms of AUROC (Area Under the ROC Curve) and anomaly scores. We use the default configuration: flow\_time\_out = 10.0s, edge\_long\_line = 15 (flows exceeding 15 packets are considered long flows), and edge\_agg\_line = 20 (threshold for short-flow aggregation). The selection of these two systems is deliberate. Exosphere represents a feature-learning supervised approach operating at the packet level, while HyperVision represents a structure-based unsupervised approach operating at the flow level. If both paradigms fail to detect TLS handshake exhaustion attacks, this suggests that the issue is systemic rather than method-specific.

Moreover, both systems have demonstrated strong performance in their original studies. Therefore, any observed detection failure is unlikely to stem from inherent system limitations, but rather from the characteristics of the attack or the shift in traffic patterns introduced by PQC.
\section{RESULTS AND ANALYSIS}

\subsection{Impact of PQC on Resource Consumption}

To answer RQ1, we investigate how PQC affects the impact of handshake exhaustion attacks, we conduct controlled experiments under identical conditions for both traditional TLS and PQC-TLS configurations. Both setups are deployed on the same AWS EC2 server with identical nginx configurations, differing only in the cryptographic algorithms used during the handshake phase. The traditional TLS setup employs standard ECDSA P-256 with X25519 key, while the PQC-TLS configuration uses ML-KEM-768 for key exchange and ML-DSA-65 for signatures. The attack parameters are kept strictly consistent across both settings. Each experiment involves 10 distributed attacker nodes launching 14 independent attack rounds, with each round generating approximately 10–15 million TLS handshake attempts. To ensure comparability, we verify that the total attack volume is nearly identical: the traditional TLS setup observes 169.9 million connection attempts (12.1M per round), while PQC-TLS observes 181.3 million (12.9M per round), with only a 6.7\% difference. Table~\ref{tab:resource_consumption} summarizes the server side resource consumption under both configurations. 

\begin{table}
\caption{Resource Consumption Under TLS Handshake Exhaustion Attack}
\label{tab:resource_consumption}
\centering
\begin{tabular}{|l|c|c|c|}
\hline
Metric & Traditional TLS & PQC-TLS & Amplification Factor \\
\hline
Peak CPU Utilization & 53.7\% & 100.0\% & 1.86$\times$ \\
\hline
Duration of CPU $>$ 50\% & 17s & 1498s & 88$\times$ \\
\hline
Duration of CPU = 100\% & 0s & 1498s & $\infty$ \\
\hline
Peak Concurrent TCP Connections & 34 & 571 & 16.8$\times$ \\
\hline
\end{tabular}
\end{table}

\begin{figure}
\centering
\includegraphics[width=\textwidth]{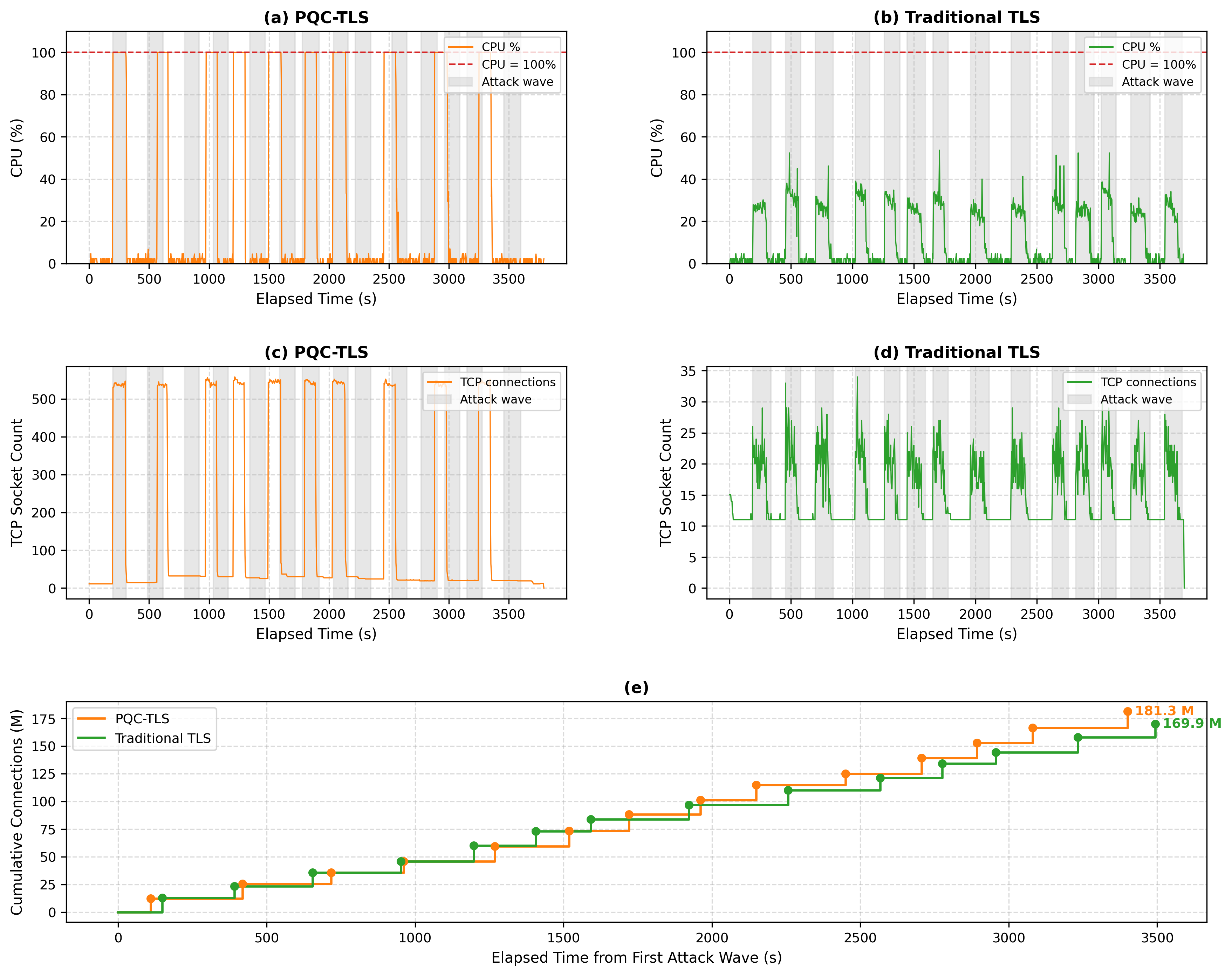}
\caption{Resource Consumption During TLS Handshake Exhaustion Attacks (a)(b) CPU utilization over time; (c)(d) TCP socket count over time; (e) cumulative connection attempts.}
\label{fig2}
\end{figure}

The main difference appears in the duration of high CPU utilization. Under traditional TLS, the server reaches a peak CPU usage of 53.7\%, and the time spent above the 50\% threshold is only 17 seconds. In contrast, under PQC-TLS, the CPU reaches full saturation (100\%) and remains at this level for 1,498 seconds, representing an 88× increase. Notably, the traditional TLS setup never approaches full CPU saturation. A similar trend is observed in TCP connection accumulation. Under PQC-TLS, the number of concurrent connections peaks at 571, with an average of 224.8, whereas traditional TLS reaches a maximum of only 34 connections. This 16.8× increase indicates that the slower handshake processing in PQC leads to significant connection backlog, as connections cannot be completed or rejected in a timely manner. A closer examination of CPU utilization over time (Fig.~\ref{fig2}) further highlights the contrast between the two configurations. 

In the traditional TLS setting, CPU usage remains below 10\% under normal traffic. When attacks occur, CPU utilization briefly spikes to around 40–50\%, but these spikes are short lived. Across the 14 attack rounds, only a few instances exceed the 50\% threshold, each lasting only a few seconds. Once an attack round ends, CPU usage quickly returns to baseline, indicating that the server can efficiently process handshake requests and release resources. In contrast, the PQC-TLS configuration exhibits a fundamentally different pattern. Even under normal traffic, the baseline CPU usage is slightly higher, around 15–20\%, reflecting the inherent computational overhead of PQC. When attacks begin, CPU utilization rapidly climbs to 100\% and remains saturated for extended periods. In most of the 14 attack rounds, the CPU stays above 90\% for several minutes. More importantly, the impact persists beyond individual attack rounds. During the intervals between attack bursts, CPU usage remains elevated at 50–70\%, indicating that the server is still processing residual handshake requests from previous attacks. This carry over effect causes resource exhaustion to accumulate over time, compounding the impact of successive attack waves and further degrading service availability. The TCP connection patterns reinforce this observation. Under traditional TLS, the number of concurrent connections remains low even during attacks, suggesting that the server can promptly complete or reject handshakes. In contrast, PQC-TLS leads to a buildup of up to 571 concurrent connections, reflecting a mismatch between connection arrival rate and processing capacity. As connections accumulate in the backlog, legitimate requests experience significant delays or even connection failures, effectively resulting in denial of service. Overall, these results indicate that, under identical attack conditions, the computational overhead introduced by PQC substantially amplifies the impact of handshake exhaustion attacks. While traditional TLS experiences only brief and recoverable load spikes (17 seconds), PQC-TLS leads to prolonged resource exhaustion lasting up to 23 minutes, pushing the system toward sustained overload.

\subsection{IDS Detection Performance}

\begin{figure}
\centering
\includegraphics[width=\textwidth]{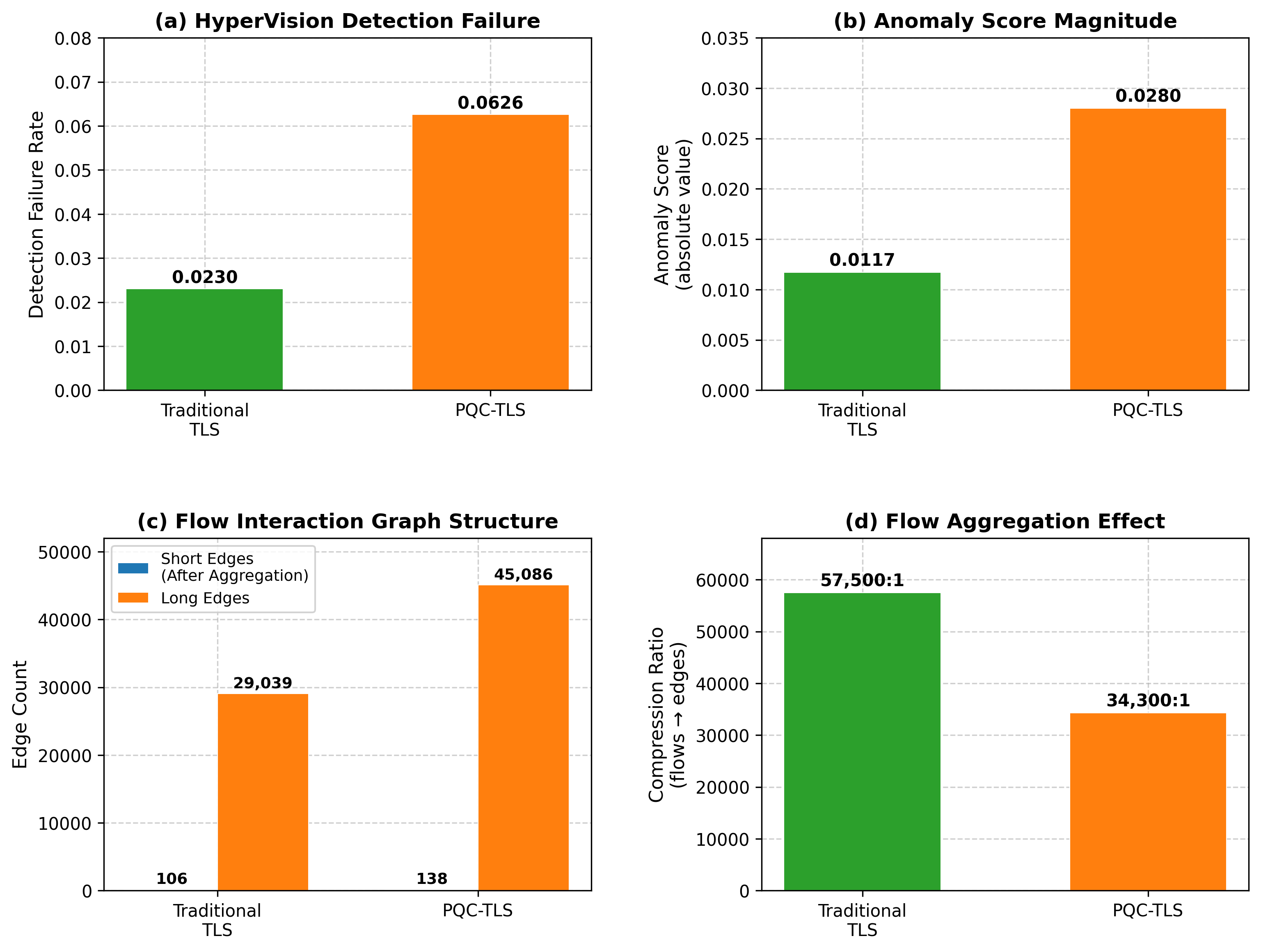}
\caption{Exosphere Detection Performance Across Baseline, Traditional TLS and PQC-TLS}
\label{fig3}
\end{figure}

To answer RQ2, we evaluate whether existing IDS can effectively detect this type of attack. We feed the generated mixed traffic datasets into the Exosphere deep learning based detection system. We use the pre-trained model provided by Exosphere without any fine-tuning or retraining for TLS handshake exhaustion, in order to evaluate its out of the box detection capability. We summarize the detection performance of Exosphere under both traditional TLS and PQC-TLS environments in the following discussion. In the traditional TLS setting, Exosphere achieves an AUC of 0.8277 and a recall of 50.18\%. Under PQC-TLS, performance further degrades, with the AUC dropping to 0.6852 and recall at 52.55\%. Although the AUC values remain slightly above the random baseline (0.5), the recall of approximately 50\% indicates that half of the attack traffic is missed, which is far from acceptable for a practical defense mechanism. In terms of precision, Exosphere achieves 75.04\% under traditional TLS and 75.65\% under PQC-TLS, suggesting that roughly three quarters of the detected packets are indeed malicious. However, when combined with the low recall, the overall F1 score ranges only from 0.34 to 0.39, reflecting severely limited detection effectiveness. It is worth noting that while Exosphere fails under both configurations, performance under PQC-TLS is consistently worse. The AUC drops by 18 percentage points (from 0.8277 to 0.6852), and the Equal Error Rate (EER) increases from 0.2162 to 0.3596, indicating degraded performance when balancing false positives and false negatives (Fig.~\ref{fig3}). 

However, since both configurations fall well below practical usability, this difference does not change the fundamental conclusion: the system fails in both cases. More importantly, if PQC specific traffic characteristics were the primary cause of detection failure, we would expect significantly better performance in the traditional TLS setting. However, this is not observed, as recall remains at only 50.18\%, and AUC is limited to 0.8277, indicating that detection is already inadequate even without PQC. To rule out implementation or configuration errors, we replicate Exosphere’s original evaluation using its benchmark dataset. The system achieves AUC = 0.9996 and F1 = 0.9913, confirming that it operates correctly under its intended conditions. This contrast highlights the severity of the issue: the same system that performs nearly perfectly (AUC > 0.99) on known attack types drops to AUC $\approx 0.83$ or lower and recall $\approx 50\%$ when faced with TLS handshake exhaustion. This suggests that the failure is not due to flaws in the IDS itself, but rather due to the distinct characteristics of handshake exhaustion attacks, which pose a fundamentally different challenge to existing detection approaches. We conduct the same evaluation using HyperVision, again using both datasets with default configurations and no modifications. The results show that HyperVision achieves an AU-ROC of 0.4943 under traditional TLS and 0.4835 under PQC-TLS both effectively equivalent to random guessing. This indicates that HyperVision is almost entirely incapable of detecting TLS handshake exhaustion attacks. Notably, HyperVision already fails in the traditional TLS setting (AU-ROC = 0.4943), and PQC only marginally worsens the result. This suggests that the failure is not primarily caused by PQC specific traffic characteristics, but instead reflects a more fundamental limitation in the detection paradigm.

\subsection{Root Cause Analysis}
To address RQ3, which focuses on the root causes of detection failure, we begin with the key observation: Exosphere fails to effectively detect TLS handshake exhaustion attacks under both traditional TLS and PQC-TLS settings, with similar levels of degradation. Based on this, we analyze several possible explanations. Hypothesis 1: PQC-induced traffic characteristics degrade detection performance. Packet-level analysis reveals notable differences between PQC-TLS and traditional TLS traffic. The average packet size in PQC-TLS is 572 bytes, approximately 2.6× larger than traditional TLS (224 bytes). In addition, large packets ($\geq 1200 bytes$) account for 21.9\% of PQC-TLS traffic, compared to only 9.7\% in traditional TLS (Table~\ref{tab:packet_distribution}). 

\begin{table}
\caption{Packet Size Distribution Comparison Between PQC-TLS and Traditional TLS}
\label{tab:packet_distribution}
\centering
\begin{tabular}{|l|c|c|}
\hline
Metric & Traditional TLS & PQC-TLS \\
\hline
Average Packet Size & 224 B & 572 B \\
\hline
Packets $>$ 1200 B (\%) & 9.7\% & 21.9\% \\
\hline
Packets $<$ 100 B (\%) & 90.2\% & 71.5\% \\
\hline
Maximum Packet Size & 1614 B & 4168 B \\
\hline
\end{tabular}
\end{table}

This hypothesis suggests that PQC-specific characteristics, such as larger certificates and longer handshake exchanges, disrupt models trained on traditional TLS traffic. Our results partially support this: detection performance under PQC-TLS is indeed worse, with AUC dropping from 0.8277 to 0.6852 (an 18 point decrease). However, PQC is not the primary cause of failure. If PQC specific features were the dominant factor, we would expect substantially better performance in the traditional TLS setting. Instead, recall remains at only 50.18\%, and AUC remains at 0.8277, indicating that detection is already inadequate even without PQC. This suggests that PQC acts more as an amplifier of existing weaknesses rather than introducing a fundamentally new problem.

Hypothesis 2: Lack of TLS layer attack samples in training data. We examine the composition of Exosphere’s training datasets, which include several widely used DDoS benchmarks. These datasets predominantly cover network and transport  layer attacks, such as UDP flooding, SYN flooding, and DNS amplification, but do not include TLS handshake exhaustion attacks. As a supervised learning system, Exosphere cannot reliably detect attack patterns it has never encountered during training. This explains the stark contrast in performance: the same system achieves near perfect results on known attack types (AUC > 0.99), yet only about 50\% recall on TLS handshake exhaustion. In this case, the failure is expected rather than surprising.

Hypothesis 3: Attack traffic is indistinguishable from benign traffic at the packet level. TLS handshake exhaustion attacks operate using legitimate TLS protocol behavior, making them highly similar to normal HTTPS traffic at both the packet and flow levels. The defining characteristics of the attack, such as high connection rates, abnormal termination patterns, and server side resource exhaustion emerge primarily at an aggregated behavioral level, rather than within individual packets or short sequences. This limitation is reflected in Exosphere’s performance. With recall at approximately 50\% and F1 scores between 0.34 and 0.39, the system struggles to extract discriminative features from packet length sequences alone. This suggests that commonly used features, such as packet size, inter arrival time, and protocol distribution, are insufficient to capture the essence of this attack. 

\begin{figure}
\centering
\includegraphics[width=\textwidth]{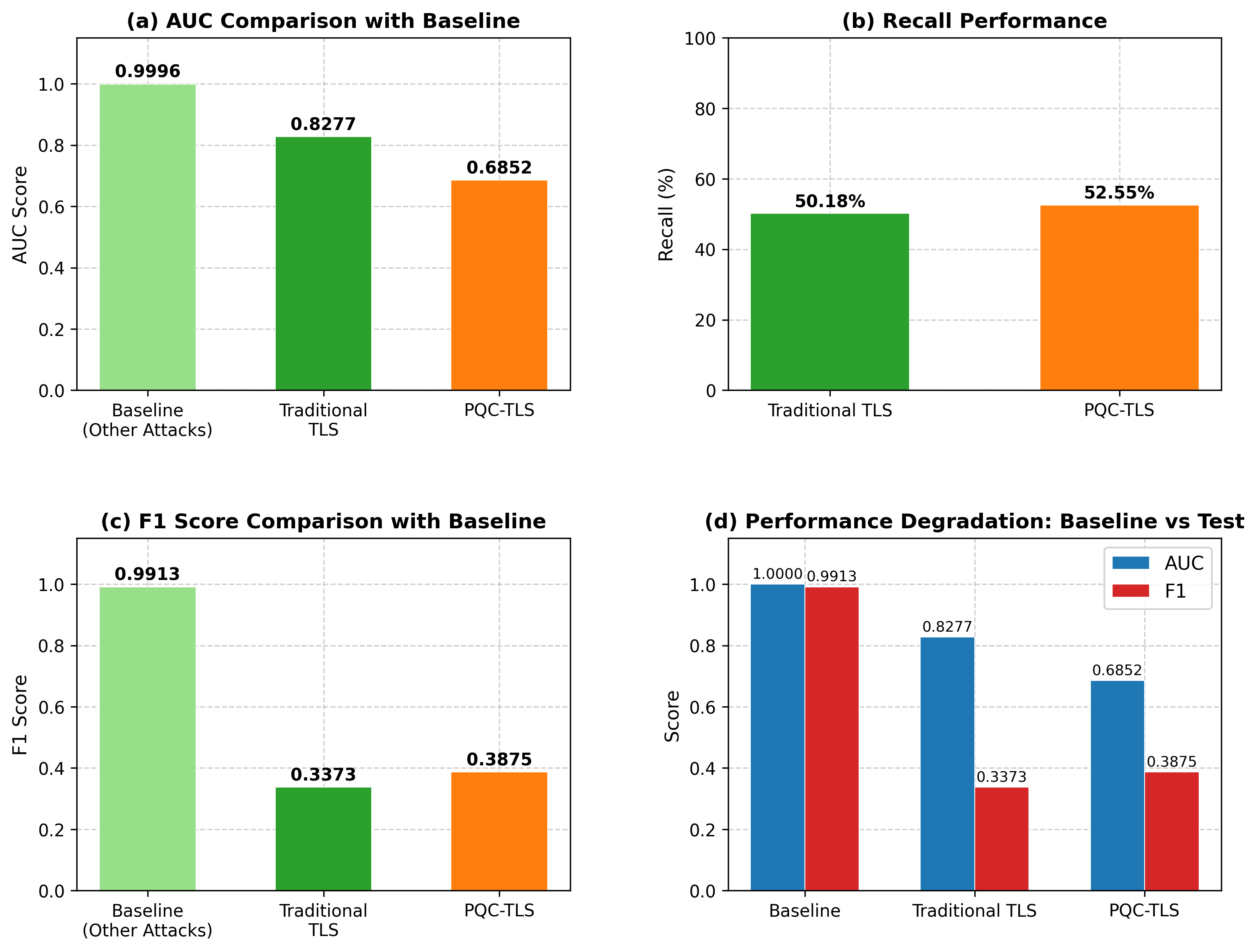}
\caption{HyperVision Structural Analysis}
\label{fig4}
\end{figure}

For HyperVision, the failure arises from a different but related issue: its short flow aggregation mechanism is structurally incompatible with this attack type. The large number of short lived connections generated by handshake exhaustion are classified as short flows (due to the 10 second timeout) and subsequently aggregated into a small number of edges. As a result, the core signal of the attack—massive connection attempts is effectively compressed away in the graph representation, leaving the model with little meaningful structure to analyze. This effect is further exacerbated in the PQC setting. The number of long edges increases by 55.3\% (from 29,039 to 45,086), as larger handshake packets cause more flows to be classified as long flows (Fig.~\ref{fig4}). 

This shift alters the graph structure and further degrades detection performance. Taken together, these findings suggest that detection failure arises from a combination of factors. First, attack traffic closely resembles benign traffic at the feature level, making it inherently difficult to distinguish using conventional features. Second, existing training datasets do not include this attack type, leaving models unprepared to recognize it. Importantly, PQC itself does not directly break detection mechanisms. Instead, it amplifies the attack’s impact, transforming what was previously considered a low priority threat into a practical and severe risk. When the computational cost of each handshake increases by several factors, and the resulting resource exhaustion is prolonged by 88×, handshake exhaustion evolves from a marginal concern into a realistic denial-of-service threat. If IDS training methodologies and feature engineering do not adapt accordingly, defense mechanisms built on outdated threat models will face an increasingly significant security blind spot.
\section{Dataset and Reproducibility}

We prepare an anonymized dataset and code package to support reproducibility during double blind review. Code and dataset will be released upon acceptance. The dataset contains two 60 minute mixed traffic PCAP traces, one for PQC-TLS and one for traditional TLS, totaling approximately 16.5 GB. Each trace preserves both benign HTTPS background traffic and TLS handshake exhaustion attack traffic without masking or sampling. Ground truth annotations are provided through attack logs that record the start time, end time, duration, and connection count of each attack round, enabling precise temporal alignment with the PCAP files.

In addition to packet traces, we provide server side telemetry collected at one second intervals, including CPU utilization, memory usage, system load, and TCP connection counts. These records support our analysis of resource amplification under PQC-TLS.

We also prepare the complete experimental framework, including AWS deployment scripts, attack generation tools, server monitoring scripts, and analysis code. The testbed consists of one EC2 server and ten distributed attack nodes, with both PQC-TLS and traditional TLS configurations deployed under the same infrastructure settings to ensure fair comparison. A detailed deployment guide, software dependencies, and implementation details are provided in the appendix. The public repository link will be released upon acceptance.
\section{DISCUSSION}
\subsection{Implications for PQC Deployment}

With the official release of post-quantum cryptographic standards by NIST, governments and enterprises are actively planning the migration toward PQC algorithms such as ML-KEM and ML-DSA. Our experimental results indicate that, while PQC algorithms are cryptographically secure, their computational characteristics may unintentionally amplify certain existing denial-of-service attack vectors. In traditional TLS environments, TLS handshake exhaustion results in relatively limited resource consumption, allowing servers to recover quickly once the attack subsides. However, when the computational cost of the handshake increases by several factors under PQC, what was previously a minor threat can escalate into a significant availability risk. This finding suggests that PQC migration should not be treated merely as a replacement of cryptographic primitives, but rather as a system wide transition requiring reevaluation of resource provisioning and threat models. Organizations should conduct capacity planning prior to deployment, ensuring that handshake processing capabilities can sustain both expected workloads and potential attack scenarios. Furthermore, conventional DDoS mitigation strategies such as rate limiting based on connection counts or traffic volume may require recalibration to reflect the increased per connection resource cost in PQC environments. Our results support a phased deployment strategy. PQC can be prioritized for internal systems and low traffic services, while high traffic public facing services may adopt hybrid modes that combine traditional and PQC algorithms, achieving a balance between quantum resistance and service availability.

\subsection{Rethinking IDS Training Paradigms}

This study reveals a fundamental limitation in current IDS training methodologies. Despite being trained on multiple widely used DDoS benchmarks and achieving near perfect performance on known attack types, Exosphere exhibits a dramatic performance drop when detecting TLS handshake exhaustion, with Recall around 50\%, effectively close to random guessing. The root cause lies in the composition of training datasets. Existing benchmarks including CIC-DDoS2019, CIC-IDS2017, Whisper datasets, and Kitsune datasets primarily focus on network layer and transport layer attacks such as UDP flood, SYN flood, and DNS amplification. None include application layer TLS attack samples, reflecting the dominant threat landscape of the past decade. However, the attack landscape is evolving. 

As infrastructure defenses improve, attackers increasingly shift toward application layer attacks that exploit the resource consumption characteristics of legitimate protocols rather than relying on volumetric flooding. TLS handshake exhaustion represents one such example; similar strategies may target HTTP/2 multiplexing, WebSocket connection establishment, or other protocol handshake phases. When training datasets fail to capture this evolution, even advanced deep learning-based IDS models develop critical blind spots. This suggests the need for a paradigm shift in dataset construction. Instead of relying on static benchmark datasets, the research community should adopt dynamic dataset update mechanisms continuously collecting emerging attack samples from real world deployments and periodically retraining or fine-tuning models. 

In addition, techniques such as few shot learning and transfer learning should be explored to enable rapid adaptation to new attack types with limited labeled data. More fundamentally, shifting toward unsupervised or semi supervised anomaly detection approaches may reduce reliance on labeled attack samples, although this introduces challenges in controlling false positive rates. Importantly, the issue extends beyond what to train on, to how features are constructed. The defining characteristics of TLS handshake exhaustion lie in server-side resource consumption and aggregated connection behaviors, rather than packet level statistical features. Current IDS systems, which primarily rely on packet or flow level features, may therefore fail to capture the essence of such attacks. Future IDS designs should consider integrating server-side telemetry, such as CPU utilization, memory usage, and connection queue length, to provide a more holistic view of system behavior.

\subsection{Limitations}

This study has several limitations. First, our experiments were conducted in a controlled cloud environment with a single server configuration and fixed attack patterns. Real world deployments may differ significantly. For example, servers may employ hardware accelerators for PQC operations, load balancers may distribute handshake processing, and application layer firewalls may enforce connection rate limits. While these factors may alter the quantitative impact of attacks, the fundamental property that PQC handshakes incur higher computational cost remains unchanged. Therefore, our core finding that PQC amplifies handshake-based attacks still holds. 

Second, we evaluated only a single model for each of two IDS systems. Although Exosphere represents a state-of-the-art deep learning-based IDS, other systems may adopt different detection paradigms or feature engineering strategies. However, the key issue identified in this work, the absence of TLS layer attack samples in training datasets is common across all major benchmarks, suggesting that our findings have broader applicability. 

Third, our attack model uses a fixed number of nodes and a predefined attack intensity. Variations in attack distribution may produce different detection characteristics. Highly distributed attacks involving thousands of nodes may be more difficult to detect using IP based aggregation features, whereas concentrated attacks from a small number of nodes may trigger simple rate limiting mechanisms. Our choice of a 10 node configuration represents a moderately distributed attack, but future work should explore the impact of attack distribution on detection performance. 

Finally, we focus on TLS handshake exhaustion as a representative application layer PQC attack. Other protocol level or PQC specific attack vectors, such as certificate validation chain abuse or post handshake authentication attacks, may exhibit different resource consumption patterns and detection challenges. This study should therefore be viewed as an initial exploration rather than a comprehensive threat assessment.

\section{CONCLUSION}

This paper investigates how post-quantum cryptography affects TLS handshake exhaustion attacks and their detectability by existing intrusion detection systems. Through an empirical evaluation in an AWS based testbed, we show that PQC substantially amplifies the resource asymmetry of the TLS handshake, turning a previously limited threat into a practical availability risk. Under the same attack intensity, PQC-TLS causes prolonged CPU saturation and a much larger TCP connection backlog than traditional TLS.

We further show that two representative IDS paradigms, Exosphere and HyperVision, both fail to effectively detect this attack. Their poor performance indicates that current IDS designs remain poorly aligned with application-layer TLS attacks, especially when attack behavior closely resembles legitimate encrypted traffic at the packet level.

Our analysis suggests that the main issue is not PQC alone, but the combination of amplified attack impact, missing TLS layer attack samples in existing benchmark datasets, and feature extraction methods that fail to capture aggregated behavioral signals and server-side resource stress. These findings suggest that PQC migration should not be treated as a purely cryptographic upgrade, but as a broader system transition requiring reevaluation of threat models, resource provisioning, and defense strategies.

Future work should focus on building PQC aware benchmark datasets, designing detection methods that incorporate both network traffic and server side telemetry, and systematically studying other protocol level attack surfaces introduced or amplified during the PQC transition.

\bibliographystyle{unsrtnat}
\bibliography{references}

\end{document}